\documentclass{article}

\usepackage{PRIMEarxiv}

\usepackage[utf8]{inputenc} 
\usepackage[T1]{fontenc}    
\usepackage{hyperref}       
\usepackage{url}            
\usepackage{booktabs}       
\usepackage{amsfonts}       
\usepackage{nicefrac}       
\usepackage{microtype}      
\usepackage{lipsum}
\usepackage{fancyhdr}       
\usepackage{graphicx}       

\usepackage{amssymb}
\usepackage{subcaption}
\usepackage{listings}
\usepackage{xcolor}

\definecolor{codegreen}{rgb}{0,0.6,0}
\definecolor{codegray}{rgb}{0.5,0.5,0.5}
\definecolor{codepurple}{rgb}{0.58,0,0.82}
\definecolor{backcolour}{rgb}{0.95,0.95,0.92}

\lstdefinestyle{mystyle}{
    backgroundcolor=\color{backcolour},   
    commentstyle=\color{codegreen},
    keywordstyle=\color{magenta},
    numberstyle=\tiny\color{codegray},
    stringstyle=\color{codepurple},
    basicstyle=\ttfamily\footnotesize,
    breakatwhitespace=false,         
    breaklines=true,                 
    captionpos=b,                    
    keepspaces=true,                 
    numbers=left,                    
    numbersep=5pt,                  
    showspaces=false,                
    showstringspaces=false,
    showtabs=false,                  
    tabsize=2
}

\title{ESAC (EQ-SANS Assisting Chatbot): Application of Large Language Models and Retrieval-Augmented Generation for Enhanced User Experience at EQ-SANS
}

\author{
  Changwoo Do, Gergely Nagy, William T. Heller \\
  Neutron Scattering Division \\
  Oak Ridge National Laboratory \\
  Oak Ridge, TN 37831\\
  \texttt{\{Changwoo Do\}doc1@ornl.gov} \\
}

\begin{document}

\begin{quote}
\vspace{10px}
\begin{center}
NOTICE OF COPYRIGHT
\end{center}
This manuscript has been authored by UT-Battelle, LLC under Contract~DE-AC05-00OR22725 with the U.S. Department of Energy (DOE). The U.S. government retains and the publisher, by accepting the article for publication, acknowledges that the US government retains a nonexclusive, paid-up, irrevocable, worldwide license to publish or reproduce the published form of this manuscript, or allow others to do so, for U.S. government purposes. DOE will provide public access to these results of federally sponsored research in accordance with the DOE Public Access Plan (http://energy.gov/downloads/doe-public-access-plan).
\end{quote}

\newpage

\maketitle

\begin{abstract}
\textit{Neutron scattering experiments have played vital roles in exploring materials properties in the past decades. While user interfaces have been improved over time, neutron scattering experiments still require specific knowledge or training by an expert due to the complexity of such advanced instrumentation and the limited number of experiments each person may perform each year. This paper introduces an innovative chatbot application that leverages Large Language Models(LLM) and Retrieval-Augmented Generation (RAG) technologies to significantly enhance the user experience at the EQ-SANS, a small-angle neutron scattering instrument at the Spallation Neutron Source of Oak Ridge National Laboratory. Through a user-centric design approach, the EQ-SANS Assisting Chatbot (ESAC) serves as an interactive reference for users, thereby facilitating the use of the instrument by visiting scientists. By bridging the gap between the users of EQ-SANS and the control systems required to perform their experiments, the ESAC sets a new standard for interactive learning and support for the scientific community using large-scale scientific facilities.}
\end{abstract}


\section{Motivation and significance}
Neutron scattering techniques have played a pivotal role in advancing materials research by providing unique insights into material structures and dynamics. Oak Ridge National Laboratory (ORNL) operates two major neutron scattering facilities: the High Flux Isotope Reactor (HFIR) and the Spallation Neutron Source (SNS).  These facilities together host over 30 neutron scattering instruments designed for studies spanning a wide range of scientific disciplines for use by visiting researchers from academia, industry and research organizations.  The detailed designs and operational principles of the instruments vary to meet the needs of the science for which they were designed to study.  Each instrument is operated in a manner dictated by the research topics that it was designed to address through control software that allows the scientist to interact with the hardware components of the instrument.  

The software that enable researchers to interact with the diverse hardware components to perform their research is a critical component of each instrument.  Examples of software frameworks used for controlling scientific instruments and experiments include CAMAC~\cite{Farren1976}, TANGO~\cite{chaize1999tango}, EPICS~\cite{Gurd2001}, and SPICE, which was built using LabVIEW~\cite{Lumsden2006}. These software frameworks not only speak directly to the instrument hardware, such as neutron detectors, sample environment equipment, and other beamline components, they efficiently process the data collected during experiments and provide the user with an interface for performing the experiment. As neutron scattering techniques have evolved, user interfaces have transitioned from manual controls, like command-line prompts, to more sophisticated software platforms that integrate experimental planning, control, and data visualization. This evolution of control interfaces has been crucial in enhancing experimental workflows and accessibility, broadening the user community, and increasing the potential for scientific discoveries using neutron scattering instruments. 

EQ-SANS (extended q-range small-angle neutron scattering) instrument at the Spallation Neutron Source has been a workhorse instrument at SNS since its commissioning in 2009\cite{Zhao2010a, Heller2018}, and information about its design and systems can be found elsewhere~\cite{Zhao2010a}. Initially, the instrument was controlled by a Python-based data acquisition system called PyDas \cite{Zolnierczuk2010} with a graphical user interface (GUI) and scripting interface developed by the instrument scientist~\cite{Zhao2010a}. In 2017, the control system software was upgraded to use the EPICS toolkit and Control System Studio (CS-Studio) \cite{Yao2021, Hatje2007CONTROLSS}. To provide a user-friendly scripting interface, a Python script wrapper was developed shortly after the transition to EPICS and CS-Studio. This front-end script interface employs Python commands that simplify activities like configuring instrument components and initiating data acquisition. Configuring the instrument at EQ-SANS involves tasks such as changing the wavelength band, moving a detector, opening or closing the secondary shutter, and adjusting the chopper frequency or phase. The simplification provided by the scripting interface improves the user experience. 
 
 Listing \ref{lst:eqsans} shows an example script for executing a series of measurements at EQ-SANS. The script begins by loading a local package containing EQ-SANS-specific functions. After setting a unique number for the experiment, known as the IPTS (Integrated Proposal Tracking System) number (line 5), various components of the instrument are configured to perform a series of transmission (the fraction of the neutron beam not absorbed by the sample) measurements at a 2.5 m sample-to-detector distance using a wavelength band with a minimum wavelength of 2.5 \AA\ at a 60 Hz chopper operation mode (line 7). Data collection begins with the command \texttt{runsampleid} (line 9) after opening the secondary shutter (line 8). In this example, transmission is measured for the empty beam, banjo cell, and porous silica sample.  The samples are mounted in one of the temperature-controlled sample environments, here a Peltier-based device, in  sample positions 1, 2, and 3, respectively. The script then runs scattering measurements (lines 14-18) after completing the transmission measurements to obtain the small-angle scattering data that will be analyzed by the researchers using the instrument.

\begin{lstlisting}[language=Python, caption=An Example EQ-SANS data collection script.]
import sys
sys.path.append("/home/controls/var/tmp/scripting/dev/")
from eqsansscanfunctions_live import *

setipts(12345)

loadconf("conf_2500mm_2p5A_60Hz_trans")
openShutter()
runsampleid("T-emptybeam 2.5m 2.5a", 0, "peltier", "pc", 1, 0.15)
runsampleid("T-banjo 2.5m 2.5a", 0, "peltier", "pc", 2, 0.15)
runsampleid("T-porsil 2.5m 2.5a", 0, "peltier", "pc", 3, 0.15)
closeShutter()

loadconf("conf_2500mm_2p5A_60Hz_scatt")
openShutter()
runsampleid("S-banjo 2.5m 2.5a", 0, "peltier", "pc", 2, 2)
runsampleid("S-porsil 2.5m 2.5a" , 0, "peltier", "pc", 3, 2)
closeShutter()
\end{lstlisting}\label{lst:eqsans}

The current Python wrapper offers a set of commands that are easier for users to remember while leveraging the flexibility of the Python language. It allows users to write experimental scripts utilizing arrays, strings, if-else statements, and for-loops, which can simplify repetitive workflows and reduce errors. However, as the number of sample environments available at the instrument increases, more functions and commands are added.  It became challenging for users to learn and remember all the commands in a short time. Although formal training is provided at the beginning of each experiment by the instrument scientists responsible for assisting users, most users visit once per year at most. Thus, providing a tool that can assist users in writing experimental scripts or obtaining help with scripting commands when the instrument scientist is not with them can significantly enhance the overall user experience and lower the technical barrier for first-time users.

To address this need, we developed the EQ-SANS Assisting Chatbot (ESAC).  ESAC not only provides descriptions and usage examples of script commands.  It is also capable of writing scripts to run experiments based on user-provided natural language descriptions of the desired measurements. This capability allows users to describe their experimental plans in simple terms, and the ESAC generates the necessary scripts automatically. Additionally, the ESAC serves as a comprehensive information resource, offering details about the instrument, safety protocols, and staff contacts. By consolidating this information, the ESAC can save users valuable time that would otherwise be spent searching through various webpages or manuals. It acts as both an information center and a virtual assistant, streamlining the experimental process and enhancing overall efficiency. The integration of this chatbot into the workflow reduces the technical barrier for new users and supports experienced researchers by providing quick access to essential information and automating script generation.

\section{Software description}
\subsection{Software architecture}

\begin{figure}[ht]
\centering\includegraphics[width=0.7\textwidth]{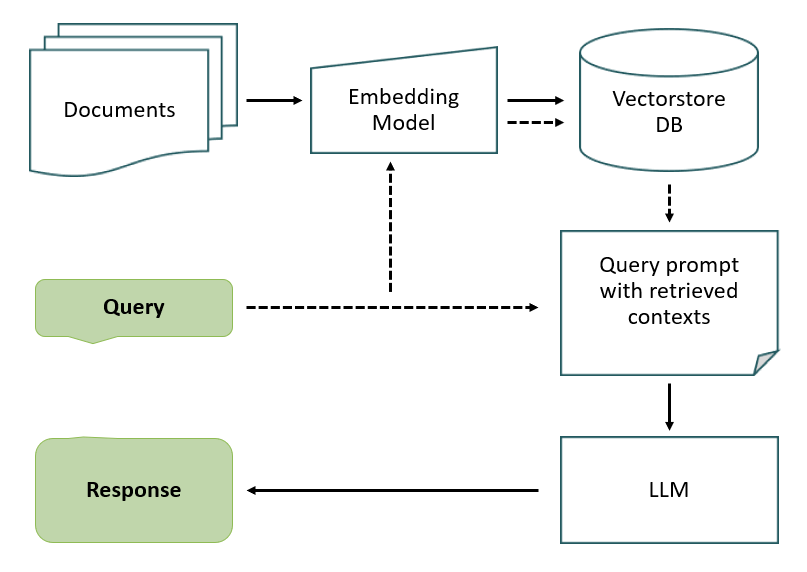}
\caption{Schematic of the RAG-LLM showing the information flow.}
\label{fig:ragllm}
\end{figure}

In order to provide information specific to EQ-SANS, we designed a chatbot  utilizing a Large Language Model (LLM), specifically chatGPT API (\url{https://openai.com}), augmented by the Retrieval-Augmented Generation (RAG) method \cite{lewis2021retrievalaugmented}. Figure \ref{fig:ragllm} illustrates the overall information flow. Initially, we established a knowledge database to enable the LLM to retrieve information based on the context and query, as demonstrated in Listing \ref{lst:code1}. We prepared knowledge documents containing detailed descriptions of experimental procedures and the functions used in Python scripts to conduct experiments. Additionally, we provided several examples of experimental scripts, complete with thorough explanations and annotations. These knowledge text files were segmented into chunks of 10,000 characters, with a 2,000-character overlap between chunks. Following this, we created a vector store using the \texttt{OpenAIEmbeddings} method, renowned for its capability to capture nuanced semantic information. The \texttt{OpenAIEmbeddings} method utilizes advanced language models from OpenAI to generate dense vector representations of text, encapsulating the semantic similarities across different textual elements. This embedding technique is well suited for our application, as it allows the chatbot to comprehend the complex details and nuances embedded within the experiment scripts and descriptions. After embedding the text chunks into numerical vectors, we used the \texttt{FAISS} package to efficiently organize and index the information, facilitating retrieval and comparison using the RAG method~\cite{lewis2021retrievalaugmented} during user interactions.

The overall information flow during user interaction is depicted in Fig. \ref{fig:ragllm}. Prior to the system's first use, the knowledge database must be established so the LLM can look for the information based on the context provided by the query. This is achieved as shown in the Listing \ref{lst:code1}. Knowledge documents were prepared by providing detailed description of experimental procedures and the functions used in the python script to run experiments. In addition, several examples of experiment scripts were also provided along with detailed description and comments. These knowledge text files were loaded and split with a chunk size of 10000 characters and a chunk overlap of 2000 characters. Subsequently, a vector store is created to store the embeddings of the text chunks using \texttt{OpenAIEmbeddings} method, known for its ability to capture nuanced semantic information. The \texttt{OpenAIEmbeddings} method leverages advanced language models, such as those developed by the OpenAI, to generate dense vector representations of text that encapsulate semantic similarities between different textual elements. This embedding method is particularly well-suited for our purposes, as it enables the chatbot to grasp the intricate details and nuances embedded within the experiment scripts and descriptions. Once the text chunks are embedded into numerical vectors, a vector store is established using \texttt{FAISS} package to efficiently organize and index the information, enabling retrieval and comparison during user interactions. 

\begin{lstlisting}[language=Python, caption=Preparing the vectorstore database.]
# Create a text splitter.
text_splitter = RecursiveCharacterTextSplitter.from_language(
    language=Language.PYTHON, chunk_size=2000, chunk_overlap=300)
texts = text_splitter.split_documents(docs)

# Create an embedding object.
embeddings = OpenAIEmbeddings()
# Create a vector store.
vectorstore = FAISS.from_documents(texts, embeddings)

# Save the vector store to disk.
vectorstore.save_local("db/vectorstore_faiss_eqsans_exp")
\end{lstlisting}\label{lst:code1}

Following the creation of the vector store, the chatbot object is constructed as shown in the Listing \ref{lst:code2}. In the code block shown, a retriever object is initialized to facilitate the retrieval of relevant information from the vector store. The retriever employs a search algorithm, specified as "mmr" (Maximal Marginal Relevance), to efficiently determine pertinent responses to user queries. The number of search items returned is set to 7 in the current example. We used \texttt{ChatOpenAI} ("gpt-4") as our language model (LLM) to provide conversational capability to our chatbot. Additionally, a memory component, termed \texttt{ConversationBufferMemory}, is introduced to store and retrieve previous chat history, enabling the chatbot to maintain continuity and coherence in conversations. By leveraging the \texttt{ConversationBufferMemory}, the chatbot can recall past interactions and responses to provide a seamless and personalized experience for users. Finally, a \texttt{ConversationalRetrievalChain} is constructed from the LLM, retriever, and memory components to form a cohesive framework for conversational retrieval. This chain integrates the functionalities of the LLM for response generation, the retriever for information retrieval, and the memory for context retention, thereby empowering the ESAC to engage in meaningful and coherent conversations about operating EQ-SANS with users.

\begin{lstlisting}[language=Python, caption=Preparing the chatbot.]
vectordb= FAISS.load_local("db/vectorstore_faiss_eqsans_exp", embeddings)
retriever = vectordb.as_retriever(search_type="mmr",
                                  search_kwargs={"k": 7})

llm = ChatOpenAI(model_name="gpt-4")
memory = ConversationBufferMemory(llm=llm, 
                                  memory_key="chat_history", 
                                  return_messages=True)
qa = ConversationalRetrievalChain.from_llm(llm, retriever=retriever, 
                                           memory=memory, verbose=False)
\end{lstlisting}\label{lst:code2}

\subsection{Preparation of knowledge database}
The knowledge documents were prepared to provide a comprehensive set of information to ESAC about performing experiments with the EQ-SANS instrument. To achieve this goal, a few principles were followed when preparing the documents. First, paragraphs were written to have clear and explicit purposes. Clear headings and subheadings delineate different topics, facilitating efficient retrieval of information by the LLM. Moreover, the use of concise and straightforward language eliminates ambiguity, ensuring that the LLM can interpret content without unnecessary complexity. Second, we tried our best to provide detailed description of the functions used in the script and the reasons for callling them. Adding more contexts to the provided knowledge documents enhances the accuracy of retrieved information. Lastly, a diverse array of examples were incorporated. This important step  turns out to be crucial for the RAG-LLM to write a full, functional script for running an experiment. With the knowledge documents provided using these principles, the ESAC can serve multiple purposes: 1) replacing traditional manuals and help files, 2) drafting a script for the experiment, and 3) searching for information about instrument, procedures, and people. The current set of knowledge documents is included in the git repository (\url{https://code.ornl.gov/ccd/esac/}).

\section{Illustrative examples}
In this section, we present illustrative examples demonstrating the multifaceted utility of ESAC as a comprehensive resource for the experiments at EQ-SANS.

\subsection{Assisting as a reference manual}
\begin{figure}[ht]
\centering\includegraphics[width=0.7\textwidth]{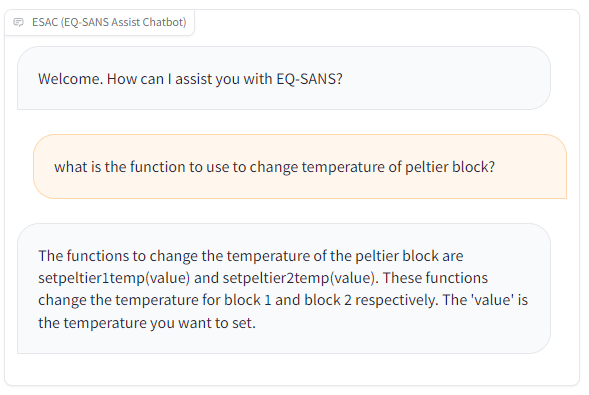}
\caption{An example query asking ESAC for a function for controlling the temperature of the Peltier-based temperature controlled sample environment.}
\label{fig:ex_peltiertemp}
\end{figure}

One of the primary functions of ESAC  is to serve as an interactive manual for the  functions and scripts used for data collection in EQ-SANS experiments. Users can inquire about specific function usage, or request information about functions based on what they do. By leveraging ESAC's knowledge base, users can easily access information about the various functional components used in experiments, thereby streamlining the experimental workflow and enhancing the user's efficiency. Fig. \ref{fig:ex_peltiertemp} shows an example of one such query. The chat interface shown in this example was created with the \texttt{Gradio} package. The example shows a user asking for a specific command to change the temperature of one of the sample environments used at EQ-SANS, the Peltier-based sample temperature controlling device.  The ESAC returns the script command for the desired task.  Although it is not shown here, it is also possible to request more specific information about the usage of such functions. 
 
\begin{figure}[ht]
\centering\includegraphics[width=0.9\textwidth]{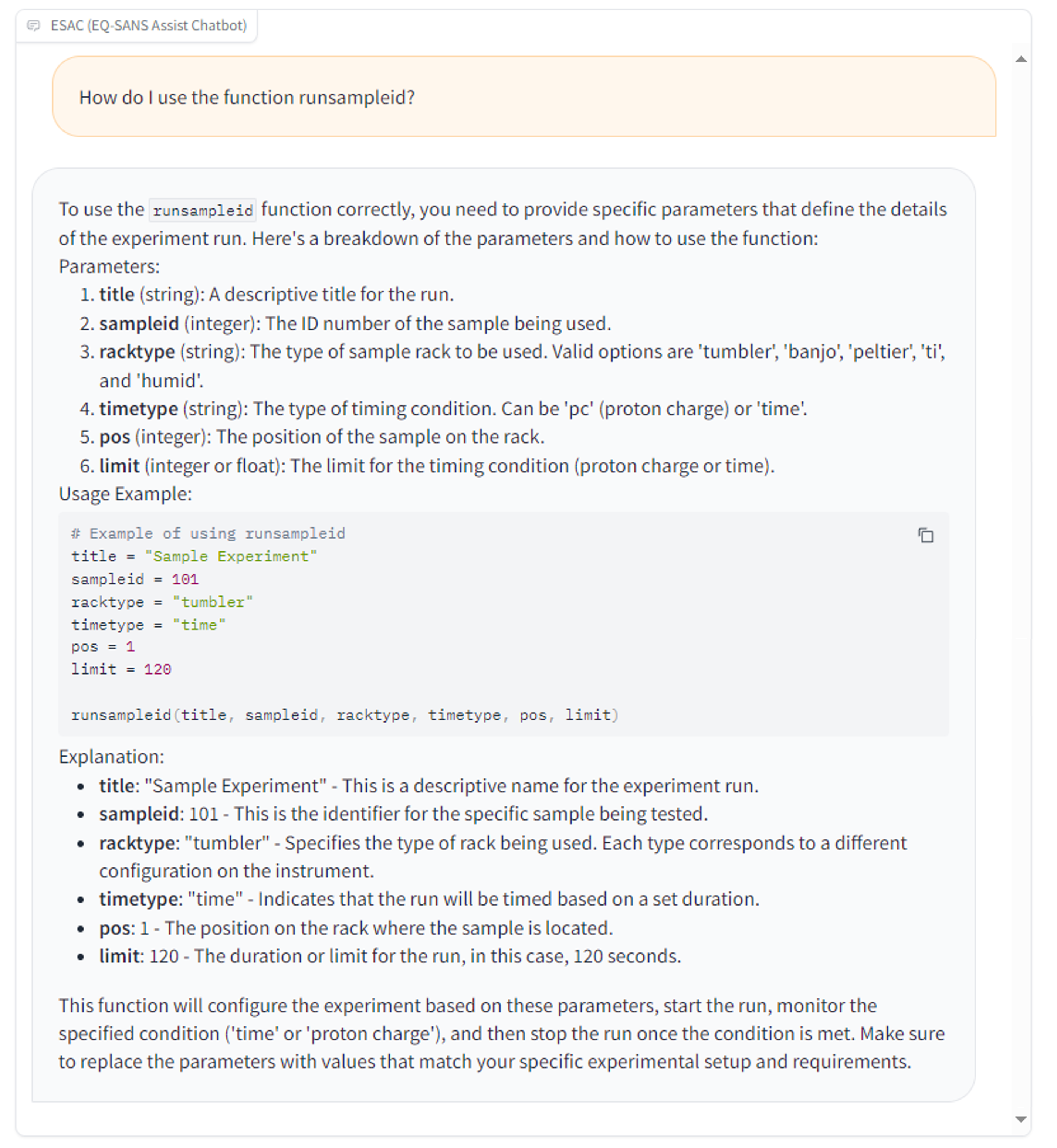}
\caption{An example query asking ESAC how to use a script function called \texttt{runsampleid}.}
\label{fig:ex_runsample}
\end{figure}

Users can also ask ESAC for more detailed function usage information. Fig. \ref{fig:ex_runsample} shows an example of for the function \texttt{runsampleid}, which is the most frequently used function at EQ-SANS.  The function tells the instrument to collect data for a sample mounted in a specific position of a sample environment until a user-specified counting limit is reached. It requires 6 input parameters. When the question was asked of the chatbot, the ESAC provided a very detailed description of the input parameters as well as an example. It is notable that the options for the rack-type parameter or the time-type were deduced from the function definition as these descriptions were not provided in the knowledge documents used to train ESAC.  

\subsection{Scripting assistant}
\begin{figure}[ht]
\centering\includegraphics[width=0.7\textwidth]{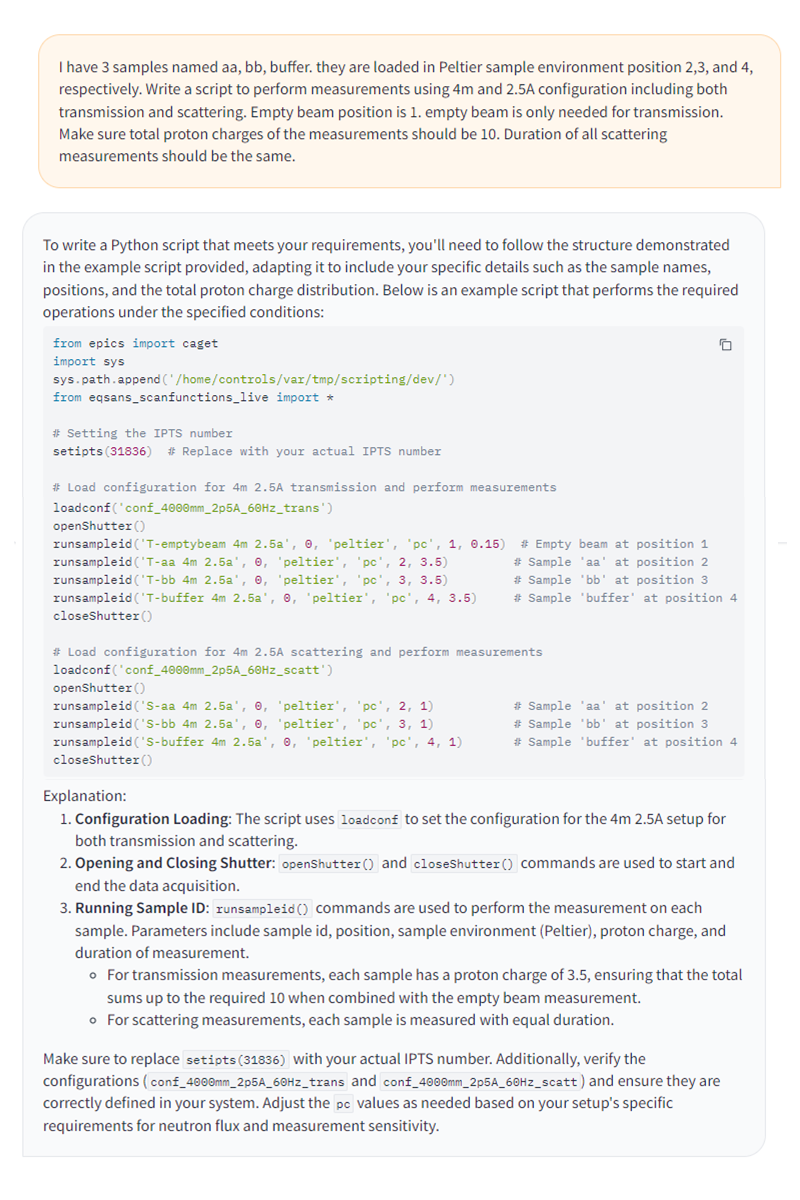}
\caption{Example query. Requesting a full script to run EQ-SANS experiment based on the sample list and desired experimental plan.}
\label{fig:ex_fullscript_a}
\end{figure}

One of the most powerful ability of the ESAC lies in the ability to generate full experiment scripts for EQ-SANS user that incorporate various experimental parameters and configurations. Users can describe the details of their experimental plans in natural English by specifying names of samples, position in the sample holders, types of sample holders, duration of measurements and the instrument configurations to be used. An example query depicted in Figure \ref{fig:ex_fullscript_a} demonstrates how users can request a full script generation from ESAC. Here, user input specified the sample names and their positions in the desired sample environment.  From the input, ESAC developed the script, which could be used as-is for data collection at EQ-SANS.  

The accuracy of ESAC is impressive, which can be partially attributed to the knowledge documents. ESAC chose to use the \texttt{runsampleid} command, and it correctly selected \texttt{peltier} for the 3rd parameter input of the command. Even though user has specified samples names only, the completed title strings (first parameter for the \texttt{runsampleid}) include ``T-'' or ``S-'' in them, indicating either a transmission or scattering measurement, respectively. In addition, as instructed in the knowledge base documents, title strings include an abbreviated form of configuration description, in this case, ``4m 2.5a'', which is helpful to have available when performing experiments using more than one instrument configuration. In the user query, the instrument configuration was given by the phrase ``4m and 2.5A configuration'', which is staff and users speak about instrument configurations because it specifies the sample-to-detector distance (4 meters) and setting for the minimum neutron wavelength (2.5 \AA). Based on this input, the chat model completed the \texttt{loadconf} command with correct configuration parameters such as \texttt{conf\_4000mm\_2p5A\_60Hz\_trans} or \texttt{conf\_4000mm\_2p5A\_60Hz\_scatt} to specify the transmission and scattering measurements, respectively. The advantage of the chat model is that it tries to analyze the user input as much as possible and reflect any requests in the generated script. 

The ESAC-generated script is not without flaws.  In this example, user made a vague request that the sum of the proton charge should be ``10'', but the request is does not specify if the sum should be the sum for all measurements or the sum for the scattering measurements. In the example shown, the chat model distributed the ``10'' proton charge among the transmission measurements. The conventions provided in the knowledge base document, the chat model assigned ``0.15'' proton charge to the emptybeam, which is a reference measurement with no sample in the beam required for calculating the neutron transmission for a sample. Then, the ESAC distributed the rest of the proton charge within the transmission measurements. It is clear what happened next. Following the next request, which is to make all scattering measurements to have the same proton charge value, chat model assigned a proton charge value of ``1'', to all the scattering measurements. While not perfect, this chat model provided a full-length script with all the right commands in a correct structure. By following the description and recommendations that are given below the full script, users should be able to make minor modifications and execute the series of measurements for their experiment. The generated script serves as an excellent template script, which is an approach to user training that the instrument staff utilize with new and returning users.  Every experiment is different, and staff work with users to determine suitable configurations and measurement times that are appropriate for the experiment performed.  

\begin{figure}[ht]
\centering\includegraphics[width=0.7\textwidth, clip=true, trim=0 40 0 0]{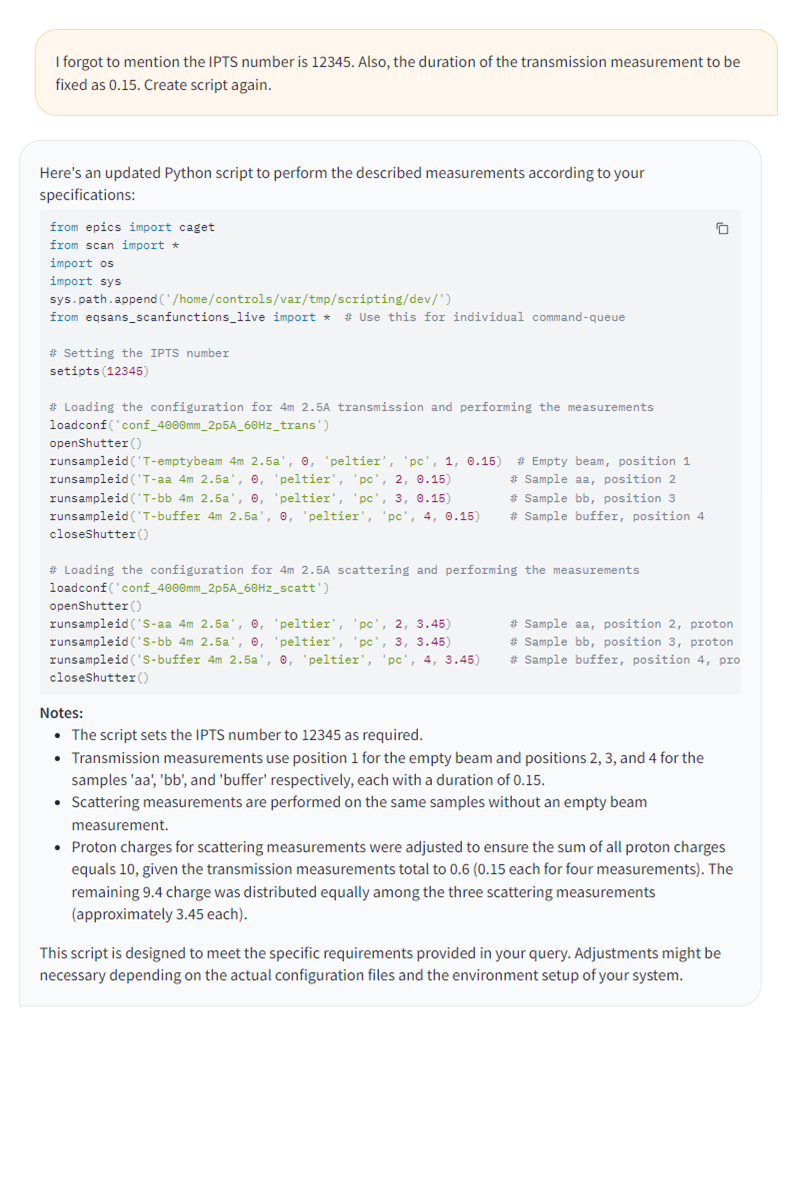}
\caption{Example query. Requesting modification of the previously generated script.}
\label{fig:ex_fullscript_b}
\end{figure}

Keeping the chat history in the current chat window enables users to interactively modify or update the scripts generated by ESAC. In Figure \ref{fig:ex_fullscript_b}, additional information, as well as correction instructions, were provided to modify the previously generated script. The requests were two fold. First, the IPTS number, which is the value goes into the command \texttt{setipts}, was updated. Second, the user requested that all transmission measurement durations be changed to ``0.15''. Both requests were successfully implemented in the new script output by ESAC. In the notes section of the output, ESAC also provided the reason behind the updated proton charge values for the scattering measurements. While the updated value of 3.45 is not one third of the 9.4 remaining after accounting for the transmission measurements, the value is sufficiently close to allow for easy adjustment.  Small adjustments to measurement duration are common, often being the final changes made before executing a data acquisition script. 

\begin{figure}[ht]
\centering\includegraphics[width=0.7\textwidth, clip=true, trim=0 10 0 0]{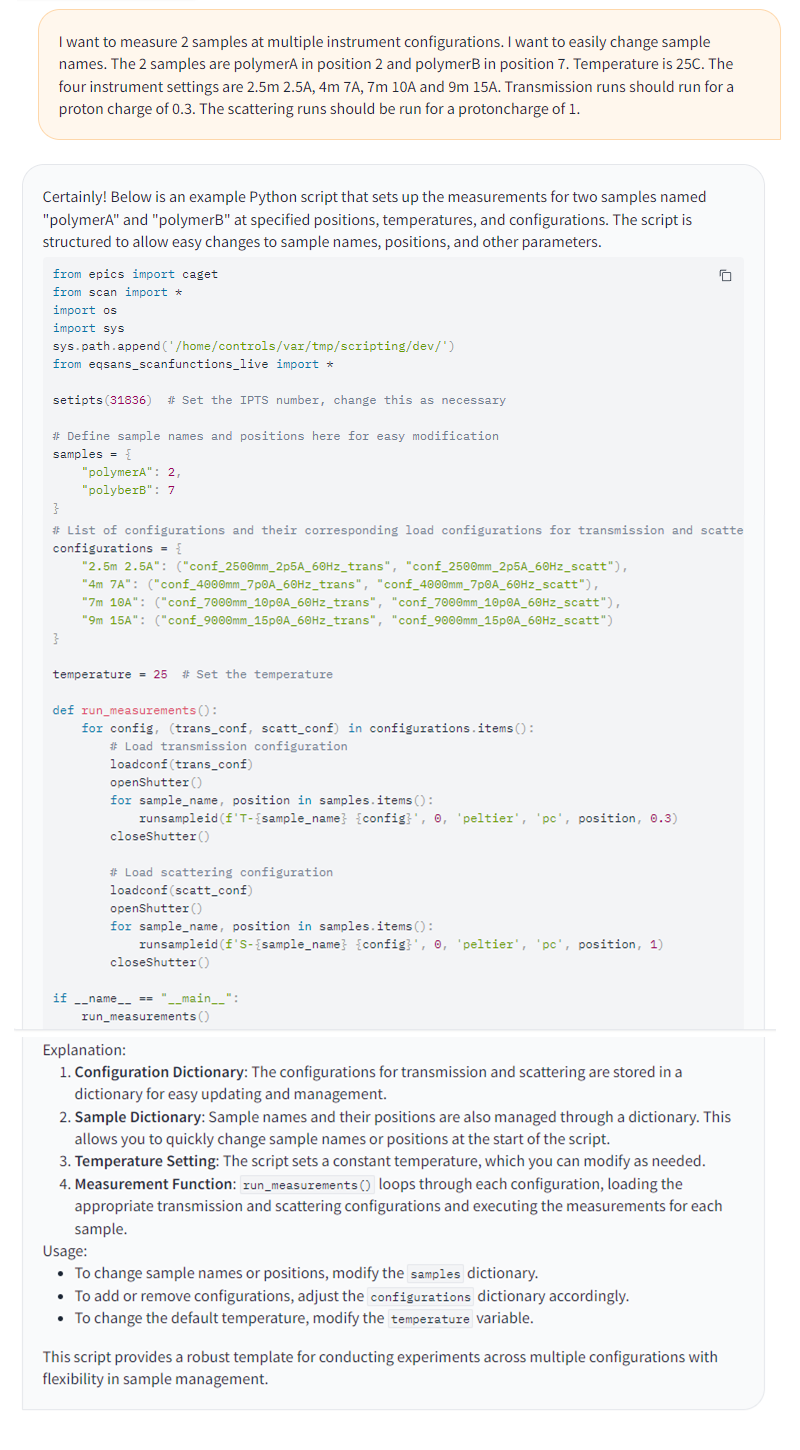}
\caption{Example query. Requesting a script in a more customizable style.}
\label{fig:ex_fullscript_c}
\end{figure}

The example scripts shown in Figures~\ref{fig:ex_fullscript_a} and~\ref{fig:ex_fullscript_b} are simple.  Most experiments are more complex, employing more than one configuration, requiring sample temperature variations, or need to use more complex sample environment equipment. ESAC can solve such problems if the user specifies the style of the script. In the example shown in Figure \ref{fig:ex_fullscript_c}, the user requests a script with more flexibility. Two key ideas incorporated was the ability to easily change sample names, and to make measurements of the samples using multiple instrument configurations. If one follows the script style from Figure~\ref{fig:ex_fullscript_a} or Figure \ref{fig:ex_fullscript_b}, the sample names must be set for every \texttt{runsampleid} command in each configuration. As a result, it can become very inconvenient to edit a more complex script, and doing so is prone to errors. The chat model recognized the example in the knowledge documents with the for-loop and arrays, and generated a script that uses a similar structure with for-loops over dictionaries containing the \texttt{configurations} and \texttt{samples}.

\begin{figure}[ht]
\centering\includegraphics[width=0.7\textwidth, clip=true, trim=0 10 0 0]{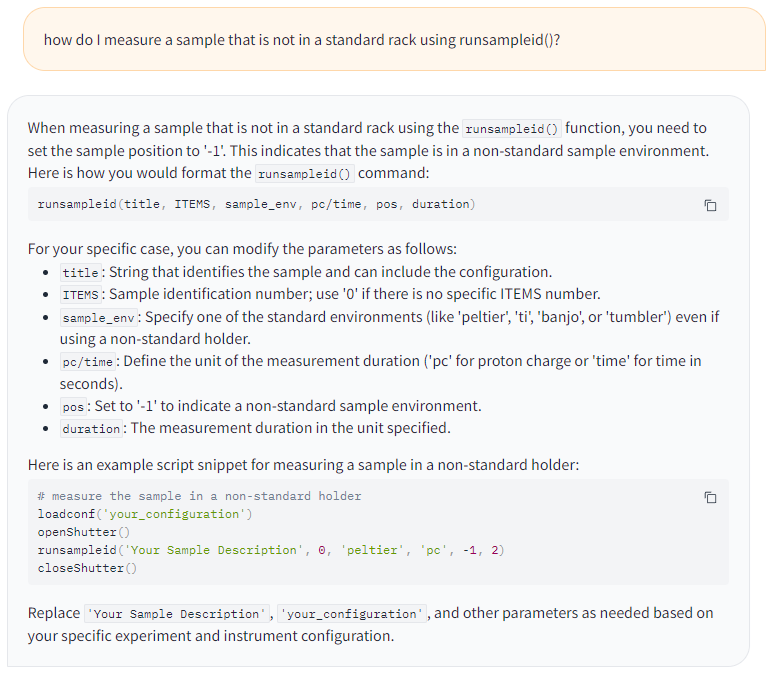}
\caption{Example query providing information for the non-standard situation.}
\label{fig:nonstandardrack}
\end{figure}

As a final test, we ensured that ESAC can provide reasonable information when a user requests information that employs a non-standard sample environment.  Non-standard sample environments include a rheometer, a tensile stage, magnets, cryostats, furnaces and the like. The same \texttt{runsampleid} command can be used to collect neutron data with sample environments that do not use the default translation stage, but the position setting that must be used does not correspond to any of the ``standard'' sample positions that the \texttt{runsampleid} would attempt to set.  For non-standard sample environments that do not have sets of defined sample positions, the position parameter needs to be set to -1, which ensures that the large translation stage used to position the sample environments does not move.  The resulting response from ESAC is provided  in Figure \ref{fig:nonstandardrack}.

\subsection{Other assistance}
During experiments, including outside regular business hours, users may have operational or safety-related inquiries. While instrument local contacts are on-call 24 hours a day, we would like to provide easy to access tools for users to search information easily regardless of when questions arise. To address this, the set of knowledge base documents provided to the ESAC includes such information. These documents enable the ESAC to also answer non-technical queries, promptly and effectively. Listings~\ref{lst:qna} and~\ref{lst:check} showcase examples of typical questions and responses, and a readiness checklist, respectively. The knowledge base encompasses crucial contact details, including those of instrument scientists. While this information is readily available on the instrument and facility websites, and it is displayed within the instrument control room, we recognize the value of streamlining its retrieval process from the instrument control computer. This approach saves users valuable time and effort, particularly during extended periods of operation. Lastly, running experiments at neutron scattering beamline can be overwhelming sometimes, particularly for new users. Experiments involve many procedures that are designed to make sure experiments can be performed safely, sometimes necessitating reaching out for assistance quickly.  

\begin{figure}[htbp]
  \begin{lstlisting}[caption={The ESAC responses to an operational query.}, label={lst:qna}]
  Q: I have a general instrument related question at 10pm, whom should I call?
  A: If you have a general question about the operation of the instrument or a suspected error at 10pm, you should contact the Instrument Hall Coordinators (IHCs). The phone number for the IHCs is 865-0241-4432.
  
  Q: I have a general instrument related question at 2pm, whom should I call?
  A: Since it is 2pm, which falls within the work hours of 8am to 6pm on weekdays, you should contact your Local Contact for any general instrument-related questions at the SNS. If you do not have the contact information for your Local Contact, you may try reaching out to the User Office at the SNS at 865-574-4600 for further assistance.
  \end{lstlisting}
  \end{figure}

  \begin{figure}[htbp]
  \begin{lstlisting}[caption={An instrument readiness checklist generated by the ESAC.}, label={lst:check}]
  Q: How do I know if I'm ready to run experiment?
  A: To determine if you are prepared to conduct a neutron scattering experiment, you can use the following checklist:
  
  1. Has the IPTS number been changed?
  2. Have the samples been checked in and assigned an ITEMS number?
  3. Is the SMA (Sample Material Agreement) signed?
  4. Is the ESS (Experiment Safety Sheet) signed?
  5. Are all users trained on the instrument?
  6. Is there a box setup for irradiated samples?
  7. Is the physical and the selected sample environment the same?
  8. Are the sample positions calibrated?
  9. Is the pinhole fully inserted?
  10. Is the beamstop adjusted at each Q setting?
  11. Is the sample table fast release fixed?
  12. Is the airflow set to the desired on or off state?
  13. Are container, empty cell, and porous silica in the rack?
  14. Is the neutron camera removed?
  
  If all these items are checked and confirmed, you are prepared to proceed with the experiment.assistance.
  \end{lstlisting}
  \end{figure}

\section{Impact}
The ESAC (EQ-SANS Assisting Chatbot) demonstrates that leveraging state-of-the-art AI can revolutionize the way neutron scattering experiments are conducted, significantly enhancing both the ease of use and accessibility of the instrumentation by providing an interactive tool that is always available to users at any time of day. By simplifying the operational complexities through an intuitive interface and streamlined command functionalities, ESAC allows researchers to concentrate on sample preparation and experiment performance rather than the intricacies of the measurement process. This user-friendly approach significantly lowers the technical barriers for entry into experiments with the instrument and makes the technique accessible to a broader range of researchers, including those without an extensive background in neutron scattering.  The tool's 24-hour availability ensures that users can obtain experimental assistance regardless of when they are working at the EQ-SANS, thus accommodating a diverse range of research schedules and increasing the overall usability of the instrument. This feature is particularly valuable in neutron scattering research, where experiments run around the clock.

ESAC has not only improved the operational efficiency and user experience at EQ-SANS but it sets a new standard for how researchers using large-scale scientific facilities interact with the complex scientific instruments there.  The broader implications for the scientific community include increased experimental throughput and a more inclusive environment where more researchers can participate in cutting-edge science. Similar opportunities for LLM integration in large user facilities have been discussed in the literature, notably in reference \cite{Prince2023}, which highlights the potential of such technologies to transform research practices. Looking to the future, ESAC can be enhanced through additional knowledge documents to provide automated assistance with more sophisticated tasks, such as the generation of complete scripts for data acquisition and data reduction after the measurement is done. Such advancements could minimize the need for active expert oversight during the experiment, streamline the experimental process, and reduce the potential for error.  
Ultimately, the ESAC will become increasingly capable of making the users of EQ-SANS more scientifically productive.  The ESAC can also be implemented for other kinds of instruments, whether they be neutron scattering instruments at ORNL, or any of the diverse range of instruments that serve researchers with a wide range of skill sets, such as those available at other neutron scattering facilities and synchrotron light sources.

\section{Conclusions}
The ESAC tool exemplifies the transformative potential of integrating advanced technologies like Large Language Models and Retrieval-Augmented Generation into scientific research environments. By bridging the gap between sophisticated experimental apparatus and user-friendliness, the ESAC not only enhances the operational efficiency of neutron scattering experiments, it also democratizes access. Readily available, automated assistance allows more members  of the scientific community to benefit from advanced research methodologies. ESAC stands as a novel innovation in scientific research technology.  As we continue to refine and expand its functionalities, it promises to elevate the standards of user interaction and experimental design at large scale user facilities.

\section*{Acknowledgements}
This research was sponsored by the Laboratory Directed Research and Development Program of Oak Ridge National Laboratory, managed by UT-Battelle, LLC, for the U.S. Department of Energy.  This research used resources at the Spallation Neutron Source, a DOE Office of Science User Facility operated by the Oak Ridge National Laboratory. The authors thank Dr. Carrie Gao for valuable contributions and ideas that significantly enhanced the usability of the developed chat model.





\bibliographystyle{unsrt}  
\bibliography{main}

\end{document}